\documentclass[reqno,a4paper]{amsart}
\usepackage{amsmath}
\usepackage{amssymb}
\usepackage{amsfonts}

\newtheorem{thm}{Theorem}

\theoremstyle{definition}

\newcommand{\be}{\begin{eqnarray}}
\newcommand{\ee}{\end{eqnarray}}

\newcommand{\edt}{elasticity difference tensor}

\begin{document}

\title{Analysing the elasticity difference tensor of general relativity}
\author{E.G.L.R.Vaz$^\ast$}
\address{$^\ast$ Departamento de Matem\'atica para a Ci\^encia e Tecnologia, Universidade do Minho\\
4800 058 Guimar\~aes, Portugal} \email{evaz@mct.uminho.pt}
\author{Irene Brito$^\natural$}
\address{$^\natural$ Departamento de Matem\'atica para a Ci\^encia e Tecnologia, Universidade do Minho\\
4800 058 Guimar\~aes, Portugal} \email{ireneb@mct.uminho.pt}

\begin{abstract}
The elasticity difference tensor, used in \cite{KS} to describe
elasticity properties of a continuous medium filling a space-time,
is here analysed from the point of view of the space-time
connection. Principal directions associated with this tensor are
compared with eigendirections of the material metric. Examples
concerning spherically symmetric and axially symmetric space-times
are then presented.
\end{abstract}

\maketitle

\section{Introduction}
In recent years there has been a growing interest in the theory of
general relativistic elasticity. Based on the classical Newtonian
elasticity theory going back to the 17th century and Hooke's law,
some authors began to adapt the theory of elasticity to the
relativity due to the necessity to study many astrophysical problems
as the interaction between the gravitational field and an elastic
solid body in the description of stellar matter, as well as to
understand the interaction of gravitational waves and gravitational
radiation and to study deformations of neutron star crusts. One of
the first elastic phenomenon considered in the relativistic context
was Weber's observation of the elastic response of an aluminium
cylinder to gravitational radiation and the detection of
gravitational waves \cite{Weber1}, \cite{Weber2} and \cite{Weber3}.
Neutron stars have attracted attention since it has been argued
\cite{Pines} that the crusts of neutron stars are in elastic states
and since it has been established the existence of a solid crust and
speculated the possibility
of solid cores in neutron stars, \cite{Shapiro}, \cite{Dermott}, \cite{Haensel}.\\
There were many attempts to formulate a relativistic version of
elasticity theory. Thereby laws of non relativistic continuum
mechanics had to be reformulated in a relativistic way. The study of
elastic media in special relativity was firstly carried out by
Noether \cite{Noether} in 1910 and by Born \cite{Born}, Herglotz
\cite{Herglotz} and Nordstr\"om \cite{Nordstrom} in 1911. The
discussion of elasticity theory in general relativity started with
Synge \cite{Synge}, De Witt \cite{Witt}, Rayner \cite{Rayner},
Bennoun \cite{B1}, \cite{B2}, Hernandez \cite{Her} and Maugin
\cite{M1} \footnote{{Relativistic elasticity has been treated in the
mid-20th century until the early seventies by many other authors.
For further references, see, for example, \cite{M1}, and for later
references see also \cite{KM92}, \cite{KS}.}}. In 1973 Carter and
Quintana \cite{CQ} developed a relativistic formulation of the
concept of a perfectly elastic solid and constructed a quasi-Hookean
perfect elasticity theory suitable for applications to high-pressure
neutron star matter. Recently, Karlovini and Samuelsson \cite{KS}
gave an important contribution to this topic, extending the results
of Carter and Quintana (see also \cite{KS2}, \cite{KS3}). Other
relevant formulations of elasticity in the framework of general
relativity were given by Kijowski and Magli (\cite{KM92},
\cite{KM97}) who presented a gauge-type theory of relativistic
elastic media and a corresponding generalization \cite{KM98}. The
same authors also studied interior solutions of the Einstein field
equations in elastic media
(\cite{M931}, \cite{M932}). \\
The recent increasing consideration of relativistic elasticity in
the literature shows the win of recognition and importance of this
topic, motivating for a detailed study of quantities used in this
context, the elasticity difference tensor defined in \cite{KS} being
one of them. This tensor occurs in the relativistic Hadamard
elasticity tensor and in the Euler equations for elastic matter.
However, one can recognize the geometric role of the elasticity
difference tensor, since, in principle, it can be used to understand
the influence of the material metric (inheriting elastic properties)
on the curvature of the space-time.

Here, in section 2, general results about relativistic elasticity
are presented. In section 3, the elasticity difference tensor is
analysed and principal directions associated with this tensor are
compared with the eigendirections of the pulled-back material
metric. A specific orthonormal tetrad is introduced to write a
general form of the elasticity difference tensor, which brings in
Ricci rotation coefficients used in the $1+3$ formalism \cite{Elst}
and the linear particle densities.

Finally, in section 4, we apply the results obtained to a static
spherically symmetric space-time and an axially symmetric
non-rotating space-time. The software \emph{Maple GRTensor} was used
to perform some calculations.

\section{General results}

Let $(M,g)$ be a space-time manifold, i.e. a $4$-dimensional,
paracompact, Hausdorff, smooth manifold endowed with a Lorentz
metric $g$ of signature $(-,+,+,+)$, $U$ being a local chart around
a point $p \in M$. Suppose that $U$ is filled with a continuum
material. The material space $\mathcal X$ is an abstract
$3$-dimensional manifold, each point in $\mathcal X$ representing an
idealized particle of the material. Moreover, the space-time
configuration of the material is described by a mapping
$$\Psi : U \subset M \longrightarrow \mathcal X ,$$
which associates to each point $p$ of the space-time the particle
$\bar p$ of the material which coincides with $p$ at a certain time.
Therefore $\Psi^{-1}(\bar p)$ represents the flowline of the
particle $\bar p$. The operators push-forward $\Psi_{\ast}$ and
pull-back $\Psi^{\ast}$ will be used  to take contravariant tensors
from $M$ to $\mathcal X$ and covariant tensors from $\mathcal X$ to
$M$, respectively, in the usual way.

If $\{\xi^{A}\}$ $(A=1,2,3)$ is a coordinate system in $\mathcal X$
and $\{\omega^{a}\}$ $(a=0,1,2,3)$ {\footnote {Capital Latin indices
A,B,... range from 1 to 3 and denote material indices. Small Latin
indices a,b,... take the values 0,1,2,3 and denote space-time
indices.}} a coordinate system in $U \subset M$, then the
configuration of the material can be described by the fields
$\xi^{A}=\xi^{A}(\omega^{a})$. The mapping $\Psi_{\ast}:
T_{p}M\longrightarrow T_{\Psi(p)} \mathcal X$ gives rise to a
$(3\times4)$ matrix (the relativistic deformation gradient) whose
entries are $\xi^{A}_{a}=\frac{\partial \xi^{A}}{\partial
\omega^{a}}$. Assuming that the world-lines of the particles
$\Psi^{-1}(\bar p)$ are timelike, the relativistic deformation
gradient is required to have maximal rank and the vector fields
$u^{a}\in T_{p}M$, satisfying $u^{a}\xi^{B}_{a}=0$, are required to
be timelike and future oriented. The vector field $u^{a}$ is the
velocity field of the matter and its components obey
$u^{a}u_{a}=-1$, $u^{a}\xi^{B}_{a}=0$ and $u^{0}>0$, \cite{KM92}.

One needs to consider, in the material space $\mathcal X$, a
Riemannian metric $\eta_{AB}$, describing the ``rest frame'' space
distances between particles calculated in the ``locally relaxed
state'' or in the ``unsheared state'' of the material and often
taken as the material metric. These approaches are presented in
\cite{KM92} and in \cite{KS}, respectively.

Let $\epsilon_{ABC}$ be the volume form of $\eta_{AB}$, with
$\Psi^{\ast} \epsilon_{ABC}= \epsilon_{abc}=\epsilon_{abcd}u^{d}$.
The particle density form is $n_{ABC}=n\epsilon_{ABC}$, with $n$ the
particle density yielding the number of particles in a volume of
$\mathcal X$ when integrated over that volume. One can define, see
\cite{KS}, a new tensor $k_{AB}$, which has $n_{ABC}$ as its volume
form and is conformal to $\eta_{AB}$:
$k_{AB}=n^{\frac{2}{3}}\eta_{AB}$. This
tensor will be taken as the {\sl material metric} in $\mathcal X$.\\
The pull-back of the material metric \be
k_{ab}=\Psi^{\ast}k_{AB}=\xi^{A}_{a}\xi^{B}_{b}k_{AB}\label{eta} \ee
and the (usual) projection tensor \be h_{ab}=g_{ab}+u_{a}u_{b} \ee
are Riemannian metric tensors on the subspace of $T_{p}M$ orthogonal
to $u^{a}$. These tensors are symmetric and satisfy
$k_{ab}u^{a}=0=h_{ab}u^{a}$.

The state of strain of the material can be measured by the
relativistic strain tensor, according to e.g. \cite{M931},
\cite{M932}: \be
s_{ab}=\frac{1}{2}(h_{ab}-\eta_{ab})=\frac{1}{2}(h_{ab}-n^{-\frac{2}{3}}k_{ab})\label{2
1}.\ee This tensor is also named as constant volume shear tensor
(see \cite{CQ}, \cite{KS}). The material is said to be ``locally
relaxed'' at a particular point of space-time if the material metric
and the projection tensor agree at that point, i.e. if the strain
tensor vanishes.

When considering elastic matter sources in general relativity, one
is confined to a stress-energy tensor taking the form $T_{ab}=-\rho
g_{ab}+2\frac{\partial \rho}{\partial g^{ab}}=\rho
u_{a}u_{b}+p_{ab}$, where $p_{ab}=2\frac{\partial\rho}{\partial
g_{ab}}-\rho h_{ab}$, the energy density being written, for
convenience, as $\rho=n\epsilon$, $\epsilon$ being the energy per
particle.

Choosing an orthonormal tetrad $\{u,x,y,z\}$ in $M$, with $u$ in the
direction of the velocity field of the matter and $x$, $y$, $z$
spacelike vectors, satisfying the orthogonality conditions $-u_a u^a
=x_a x^a = y_a y^a = z_a z^a = 1$, all other inner products being
zero, the space-time metric can be written as \be
g_{ab}=-u_{a}u_{b}+h_{ab}=-u_{a}u_{b}+x_{a}x_{b}+y_{a}y_{b}+z_{a}z_{b}.
\label{7}\ee Here we will choose the spacelike vectors of the tetrad
along the eigendirections of $k^a_b=g^{ac}k_{cb}$, so that \be
k_{ab}=n_{1}^{2}\hspace{0.6mm}x_{a}x_{b}+n_{2}^{2}\hspace{0.6mm}y_{a}y_{b}+n_{3}^{2}\hspace{0.6mm}z_{a}z_{b},
\label{6}\ee where $n_{1}^{2}$, $n_{2}^{2}$ and $n_{3}^{2}$ are the
(positive) eigenvalues of $k_{a}^{b}$. The linear particle densities
$n_1$, $n_2$ and $n_3$ satisfy $n=n_1n_2n_3$. It should be noticed
that those eigenvectors are automatically orthogonal whenever the
eigenvalues referred above are distinct. However, if the eigenvalues
are not all distinct, the eigendirections associated to the same
eigenvalue can (and will) be chosen orthogonal.

It is convenient to consider the spatially projected connection
$D_a$ acting on an arbitrary tensor field $t^{b...}_{c...}$ as
follows: \be
D_{a}t^{b...}_{c...}=h^{d}_{a}h^{b}_{e}...h^{f}_{c}...\nabla_{d}t^{e...}_{f...}.\label{8}\ee
Here $\nabla$ is the connection associated with $g$ and one has $D_a
h_{bc}=0$. Another operator $\tilde{D}$, such that its action on the
same tensor is \be
\tilde{D}_{a}t^{b...}_{c...}=h^{d}_{a}h^{b}_{e}...h^{f}_{c}...\tilde{\nabla}_{d}t^{e...}_{f...}\label{9}
\ee is also considered. One has \be \tilde{D}_b X^a = D_b X^a +
S^a_{\hspace{0.15cm}bc} X^c, \label{10} \ee for any space-time
vector field $X$. The tensor field $S^a_{\hspace{0.15cm}bc}$ is the
elasticity difference tensor introduced by Karlovini and Samuelsson
in \cite{KS}.  This third order tensor can be written as \be
S^{a}_{\hspace{0.15cm}bc}=\frac{1}{2}k^{-am}(D_{b}k_{mc}+D_{c}k_{mb}-D_{m}k_{bc}),
\label{11}\ee where $k^{-am}$ is such that
$k^{-am}k_{mb}=h^{a}_{b}$. This tensor is used by the same authors
to write the Hadamard elasticty tensor, used to describe elasticity
properties in space-time, and the Euler equations $\nabla_b
T^{ab}=0$ for elastic matter.

The covariant derivative of the timelike unit vector field $u$ can
be decomposed as follows \be u_{a;b}=-\dot{u}_a u_b+ D_b
u_a=-\dot{u}_a u_b+\frac{1}{3} \Theta
h_{ab}+\sigma_{ab}+\omega_{ab},\label{covu} \ee where
$\dot{u}_{\alpha}$ is the acceleration, $\sigma_{\alpha\beta}$, the
symmetric tracefree rate of shear tensor field,
$\omega_{\alpha\beta}$, the antisymmetric vorticity tensor field and
$\Theta$, the expansion scalar field for the congruence associated
with $u$.

\section{Properties of the Elasticity Difference Tensor}

Here we will investigate the algebraic properties of the \edt. This
tensor, important when studying elasticity within the framework of
general relativity, is related to the connection of the space-time,
as shown in the previous section. The following two properties of
the elasticity difference tensor are straightforward:
\begin{enumerate}
\item[(i)] it is symmetric in the two covariant indices, i. e. \be
S^{a}_{\hspace{0.15cm}bc}=S^{a}_{\hspace{0.15cm}cb}; \ee
\item[(ii)]it is a completely flowline orthogonal tensor field,
i.e. \be
S^{a}_{\hspace{0.15cm}bc}u_{a}=0=S^{a}_{\hspace{0.15cm}bc}u^{b}=S^{a}_{\hspace{0.15cm}bc}u^{c}.
\ee
\end{enumerate}
The elasticity difference tensor can be approached using the
space-time connection, as will be shown here.

It is a well known result that the difference between two
connections $\tilde{\nabla}$ and $\nabla$, associated with two
different metrics $\tilde{g}$ and $g$, respectively, defined on $U$,
is the following $(1,2)$ tensor: \be
C^{n}_{\hspace{0.15cm}ml}=\tilde{\Gamma}^{n}_{\hspace{0.15cm}ml}-\Gamma^{n}_{\hspace{0.15cm}ml}\label{01},\ee
$\tilde{\Gamma}^{n}_{\hspace{0.15cm}ml}$ and
$\Gamma^{n}_{\hspace{0.15cm}ml}$ being the Christoffel symbols
associated with those two metrics. In a local chart, this tensor can
be written as (\cite{W}, \cite{M}) \be
C^{n}_{\hspace{0.15cm}ml}=\frac{1}{2}\tilde{g}^{np}(\tilde{g}_{pm;l}+\tilde{g}_{pl;m}-\tilde{g}_{ml;p})\label{02},\ee
where $\tilde{g}^{np}$ is such that
$\tilde{g}^{np}\tilde{g}_{pr}=\delta^{n}_{r}$ and  a semi-colon $;$
represents the covariant derivative with respect to $g$. The
difference tensor $C^{n}_{\hspace{0.15cm}ml}$ can be used to write
the difference of the Riemann and the Ricci tensors associated with
the two metrics in the following form (see e.g. \cite{R}): \be
\tilde{R}^{a}_{\hspace{0.15cm}bcd}-R^{a}_{\hspace{0.15cm}bcd}=-C^{a}_{\hspace{0.15cm}bd;c}+
C^{a}_{\hspace{0.15cm}bc;d}-C^{a}_{\hspace{0.15cm}lc}C^{l}_{\hspace{0.15cm}bd}+
C^{a}_{\hspace{0.15cm}ld}C^{l}_{\hspace{0.15cm}bc}\label{riemann}
\ee and \be
\tilde{R}_{bd}-R_{bd}=-C^{a}_{\hspace{0.15cm}bd;a}+C^{a}_{\hspace{0.15cm}ba;d}-
C^{a}_{\hspace{0.15cm}la}C^{l}_{\hspace{0.15cm}bd}+C^{a}_{\hspace{0.15cm}ld}C^{l}_{\hspace{0.15cm}ba}\label{ricci}.
\ee The projection of the difference tensor orthogonally to $u$ is
defined by the expression \be
h^{a}_{n}\hspace{1mm}h^{m}_{b}\hspace{1mm}h^{l}_{c}\hspace{1mm}C^{n}_{\hspace{0.15cm}ml}\label{05}.
\ee When the connections used to define the difference tensor are
associated with the metrics $g_{ab}=- u_a u_b+h_{ab}$ and $
\tilde{g}_{ab}=- u_au_b+k_{ab}$, then the corresponding difference
tensor, projected according to \eqref{05}, yields \eqref{11} i.e.
the {\sl{elasticity difference tensor}} defined in the previous
section.

Under this approach, the \edt \ is the projection, orthogonal to
$u$, of the difference between two connections, one associated with
the space-time metric and the other with the metric
$\tilde{g}_{ab}=- u_au_b+k_{ab}$, where $k_{ab}$ is the pull-back of
the material metric $k_{AB}$.

Calculating the spatially projected versions of equation
\eqref{riemann}, using \eqref{8} and \eqref{05}, yields the
following expression for the difference of the Riemann tensors: \be h_{m}^{f}\,h_{g}^{n}\,h_{e}^{p}\,h_{h}^{q}\,[\,h_{a}^{m}\,h_{n}^{b}\,h_{p}^{c}\,h_{q}^{d}\,(\tilde{R}^{a}_{\hspace{0.15cm}bcd}-R^{a}_{\hspace{0.15cm}bcd})]\nonumber\\
=-D_{e}S^{f}_{\hspace{0.15cm}gh}+D_{h}S^{f}_{\hspace{0.15cm}ge}-S^{f}_{\hspace{0.15cm}ke}S^{k}_{\hspace{0.15cm}gh}+S^{f}_{\hspace{0.15cm}kh}S^{k}_{\hspace{0.15cm}ge}.\ee

The spatially projection of \eqref{ricci}, the difference of the
Ricci tensors, can be obtained analogously by equating the indices
$a=c$ in the last expression.

Therefore, these expressions, which contain the elasticity
difference tensor, give the difference between the Riemann and Ricci
tensors associated with the metrics referred to above.

Now we will obtain the tetrad components of the elasticity
difference tensor. To do so, it is more convenient to use the
following notation for the orthonormal tetrad:
$e_{\mu}^{a}=(e_{0}^{a},e_{1}^{a},e_{2}^{a},e_{3}^{a})=(u^{a},x^{a},y^{a},z^{a}).$
Tetrad indices will be represented by greek letters from the second
half or the first half of the alphabet according to their variation
as follows: $\mu,\nu,\rho...=0-3$ and $\alpha,\beta,\gamma...=1-3$.
The Einstein summation convention and the notation for the symmetric
part of tensors will only be applied to coordinate indices, unless
otherwise stated. The operation of raising and lowering tetrad
indices will be performed with
$\eta_{\mu\nu}=\eta^{\mu\nu}=diag(-1,1,1,1)$ and one has
$g_{ab}=e_{\mu a}e_{\nu b}\eta^{\mu\nu}.$

Writing the Ricci rotation coefficients  as
$\gamma_{\mu\nu\rho}=e_{\mu a;b}e^{a}_{\nu}e^{b}_{\rho}$, the tetrad
components of the elasticity difference tensor can be obtained using
the standard relationship \be
S^{\alpha}_{\hspace{0.15cm}\beta\gamma}=S^{a}_{\hspace{0.15cm}bc}e_{a}^{\alpha}e^{b}_{\beta}e^{c}_{\gamma}\ee
the result being \be \label{S1} \begin{split}
S^{\alpha}_{\hspace{0.15cm}\beta\gamma}&=\frac{1}{2n_{\alpha}^{2}}[\left(n_{\alpha}^{2}-n^{2}_{\gamma}\right)
\gamma^{\alpha}_{\hspace{0.15cm}\gamma\beta}+
\left(n_{\alpha}^{2}-n_{\beta}^{2}\right)\gamma^{\alpha}_{\hspace{0.15cm}\beta\gamma}+
\left(n_{\gamma}^{2}-n^{2}_{\beta}\right)\gamma_{\beta\gamma}^{\hspace{0.3cm}\alpha}+
D_{n}(n^{2}_{\alpha})e^{n}_{\beta}\delta^{\alpha}_{\gamma}\\
&
+D_{p}(n^{2}_{\alpha})e^{p}_{\gamma}\delta^{\alpha}_{\beta}-D_{l}(n^{2}_{\beta})e^{l\alpha}\delta_{\beta\gamma}].
\end{split}
\ee An alternative form for the last expression is: \be \label{S2}
\begin{split}
S^{\alpha}_{\hspace{0.15cm}\beta\gamma}&=\frac{1}{2}[(1-\epsilon_{\gamma\alpha})\gamma^{\alpha}_{\hspace{0.15cm}\gamma\beta}+
(1-\epsilon_{\beta\alpha})\gamma^{\alpha}_{\hspace{0.15cm}\beta\gamma}+
(\epsilon_{\gamma\alpha}-\epsilon_{\beta\alpha})\gamma_{\beta\gamma}^{\hspace{0.3cm}\alpha}+m_{\beta\alpha}\delta^{\alpha}_{\gamma}+ m_{\gamma\alpha}\delta^{\alpha}_{\beta}\\
&
-m^{\alpha}_{\hspace{0.15cm}\beta}\delta_{\beta\gamma}\epsilon_{\beta\alpha}],
\end{split}
\ee where
$\epsilon_{\gamma\alpha}=\left(\displaystyle{\frac{n^{2}_{\gamma}}{n^{2}_{\alpha}}}\right)$
and $m^{\alpha}_{\hspace{0.15cm}\beta}=D_{a}(\ln
n^{2}_{\beta})e^{a\alpha}.$

The Ricci rotation coefficients, when related to the quantities used
in the decomposition \eqref{covu}, can be split into the set
\cite{EE}:
\begin{align}
\gamma_{0 \alpha 0}&= \dot{u}_{\alpha}\label{rrc1}\\
\gamma_{0 \alpha \beta}&= \frac{1}{3}\Theta \delta_{\alpha \beta} +
\sigma_{\alpha \beta}-\epsilon_{\alpha \beta \gamma }
\omega^{\gamma} \label{rrc2}\\
\gamma_{\alpha \beta 0}&= -\epsilon_{\alpha \beta \gamma }
\Omega^{\gamma} \label{rrc3}\\
\gamma_{\alpha \beta \gamma}& = -A_{\alpha}\delta_{\beta
\gamma}+A_{\beta}\delta_{\alpha\gamma} -
\frac{1}{2}(\epsilon_{\gamma \delta\alpha}
N^{\delta}_{\hspace{0.15cm}\beta}
-\epsilon_{\gamma\delta\beta}N^{\delta}_{\hspace{0.15cm}\alpha}+
\epsilon_{\alpha \beta \delta } N^{\delta}_{\hspace{0.15cm}\gamma
}).\label{rrc4}
\end{align}

The quantities $A$ and $N$ appear in the decomposition of the
spatial commutation functions $\Gamma^{\alpha}_{\hspace{0.15cm}\beta
\gamma} = \gamma^{\alpha}_{\hspace{0.15cm}\gamma \beta} -
\gamma^{\alpha}_{\hspace{0.15cm}\beta \gamma}$, given in \cite{EM},
where $N$ is a symmetric object.

The \edt\ can be expressed using three second order symmetric
tensors, here designated as $
\underset{\hspace{-0.3cm}\alpha}{M_{bc}}$, $\alpha = 1,2,3$,  as
follows: \be
S^a_{\hspace{0.15cm}bc}=\underset{\hspace{-0.3cm}1}{M_{bc}}x^a+\underset{\hspace{-0.3cm}2}{M_{bc}}y^a+\underset{\hspace{-0.3cm}3}{M_{bc}}z^a
= \sum_{\alpha =1}^3 \underset{\hspace{-0.3cm}\alpha}{M_{bc}}
e_{\alpha}^a. \ee Here we will study some properties of the three
tensors $\underset{\hspace{-0.3cm}\alpha}{M_{bc}}$ in order to
understand until which extent the principal directions of the pulled
back material metric remain privileged directions of the \edt, i.e.
of the tensors $\underset{\hspace{-0.3cm}\alpha}{M_{bc}}$, by
studying the eigenvalue-eigenvector problem for these second order
tensors.

First, we will obtain a general expression for
$\underset{\hspace{-0.3cm}\alpha}{M_{bc}}$, $\alpha = 1,2,3$, which
depends explicitely on the orthonormal tetrad vectors, the Ricci
rotation coefficients and the linear particle densities
$n_{\alpha}$. This comes from the contraction of
$S^a_{\hspace{0.15cm}bc}$ in \eqref{11} with each one of the spatial
tetrad vectors, followed by the use of the relationships \eqref{6},
\eqref{8} and appropriate simplifications. The final result is \be
\label{M} \begin{split}
\underset{\hspace{-0.3cm}\alpha}{M_{bc}}&=u^{m}(e_{\alpha
m;(b}u_{c)}+u_{(b}e_{\alpha
c);m})+e_{\alpha(b;c)}-e_{\alpha}^{m}e_{\alpha(c}e_{\alpha b);m}\\
&+\gamma_{0\alpha\alpha} \ u_{(b}e_{\alpha c)}- \
\gamma_{0\alpha0} \ u_{b}u_{c}\\
&+\frac{1}{n_{\alpha}}[2n_{\alpha,(b}e_{\alpha
c)}+2n_{\alpha,m}u^{m}u_{(b}e_{\alpha
c)}+n_{\alpha,m}e_{\alpha}^{m}e_{\alpha b}e_{\alpha c}]\\
&+\frac{1}{n_{\alpha}^{2}}\{-e_{\alpha}^{m}(e_{\beta b}e_{\beta
c}n_{\beta}n_{\beta,m}+e_{\gamma b}e_{\gamma
c}n_{\gamma}n_{\gamma,m})\\
&+n_{\gamma}^{2}[(\gamma_{0\gamma\alpha}-\gamma_{\alpha\gamma
0})u_{(b}e_{\gamma c)}+e_{\alpha}^{m}(e_{\gamma m;(b}e_{\gamma
c)}-e_{\gamma(b}e_{\gamma c);m})]\\
&+n_{\beta}^{2}[(\gamma_{0\beta\alpha}-\gamma_{\alpha\beta
0})u_{(b}e_{\beta c)}+e_{\alpha}^{m}(e_{\beta m;(b}e_{\beta
c)}-e_{\beta(b}e_{\beta c);m})]\},
\end{split}
\ee

where $\gamma\neq\beta\neq\alpha$, for one pair $(\beta,\gamma)$, a
comma being used for partial derivatives. It should be noticed that
this expression also contains the non-spatial Ricci rotation
coefficients given in \eqref{rrc1}, \eqref{rrc2} and \eqref{rrc3}.

Naturally, the expressions obtained for
$\underset{\hspace{-0.3cm}\alpha}{M_{bc}}$ still satisfy the
conditions $\underset{\hspace{-0.3cm}\alpha}{M_{bc}} u^b=0$.

The eigenvalue-eigenvector problem for
$\underset{\hspace{-0.3cm}\alpha}{M_{bc}}$ is quite difficult to
solve in general. However, one can investigate the conditions for
the tetrad vectors to be eigenvectors of those tensors, the results
being summarized in the two following theorems.

Intrinsic derivatives of arbitrary scalar fields $\Phi$, as
derivatives along tetrad vectors, will be represented
$\Delta_{e_{\alpha}}$ and defined as:
$$ \Delta_{e_{\alpha} }\Phi = \Phi_{,m} e_{\alpha}^m , $$
where a comma is is used for  partial derivatives.
\begin{thm} The tetrad vector $e_{\alpha}$ is an eigenvector for $\underset{\alpha}{M}$ iff
$n_{\alpha}$ remains invariant along the two spatial tetrad vectors
$e_{\beta}$, such that $ \beta \neq \alpha $ i.e.
$\Delta_{e_{\beta}}(\ln n_{\alpha})= 0$ whenever $\beta \neq \alpha
.$

The corresponding eigenvalue is $\lambda=\Delta_{e_{\alpha}}(\ln
n_{\alpha})$. \label{theorem2}\end{thm}

\noindent {\it Proof:} In order to solve this eigenvector-eigenvalue
equation the following algebraic conditions are used \be
\underset{\hspace{-0.3cm}\alpha}{M_{b}^{c}}e_{\alpha}^{b}e_{\alpha
c}=\lambda \label{a11}, \ee \be
\underset{\hspace{-0.3cm}\alpha}{M_{b}^{c}}e_{\alpha}^{b}e_{\beta
c}=0 \label{a12}\ee and \be
\underset{\hspace{-0.3cm}\alpha}{M_{b}^{c}}e_{\alpha}^{b}e_{\gamma
c}=0 \label{a13}, \ee where $\gamma\neq\beta\neq\alpha$. Using the
orthogonality conditions satisfied by the tetrad vectors and the
properties of the rotation coefficients, namely the fact that they
are anti-symmetric on the first pair of indices, (\ref{a12}) and
(\ref{a13}) yield $\Delta_{e_{\beta}}(\ln
n_{\alpha})=0=\Delta_{e_{\gamma}}(\ln n_{\alpha})$ so that
$\Delta_{e_{\beta}}n_{\alpha}=0=\Delta_{e_{\gamma}}n_{\alpha}$. On
the other hand from (\ref{a11}) one obtains
$\lambda=\Delta_{e_{\alpha}}(\ln n_{\alpha})$.\\
It should be noticed that $\lambda=0$ whenever $n_{\alpha}$ remains
constant along $e_{\alpha}$. However this condition is equivalent to
$n_{\alpha}=c$, with $c$ a constant. In this case,
$k_{ab}=c^{2}\hspace{0.6mm}e_{\alpha a}e_{\alpha
b}+\underset{\beta\neq \alpha}{\sum}n^{2}_{\beta}e_{\beta a}e_{\beta
b}$. \hfill $\square$

\begin{thm} $e_{\beta}$ is an eigenvector of $\underset{\alpha}{M}$ iff  the following conditions are satisfied:
 \begin{enumerate}
 \item[ \rm{(i)}]
  $\Delta_{e_{\beta}}(\ln n_{\alpha})=0$, i.e. $n_{\alpha}$ remains
invariant along the direction of $e_{\beta}$;

 \item[ \rm{(ii)}]$\gamma_{\alpha\gamma\beta}[n_{\alpha}^{2}-n_{\gamma}^{2}]+
\gamma_{\alpha\beta\gamma}[n_{\alpha}^{2}-n_{\beta}^{2}]+
\gamma_{\beta\gamma\alpha}[n_{\gamma}^{2}-n_{\beta}^{2}]=0$, where
$\gamma\neq\beta\neq\alpha$ for one pair $(\beta,\gamma)$.
\end{enumerate}
The corresponding eigenvalue is
$\lambda=-\frac{n_{\beta}}{n_{\alpha}^{2}}\Delta_{e_{\alpha}}n_{\beta}+
\gamma_{\alpha\beta\beta}(-\frac{n_{\beta}^{2}}{n_{\alpha}^{2}}+1)$.
\label{theorem3}\end{thm}

\noindent {\it Proof:} Contracting
$\underset{\hspace{-0.3cm}\alpha}{M_{b}^{c}}e_{\beta}^{b}=\lambda
e_{\beta}^{c}$ with $e_{\alpha c}$ one obtains
$\Delta_{e_{\beta}}(\ln n_{\alpha})=0$. This condition is satisfied
whenever $\Delta_{e_{\beta}}n_{\alpha}=0$. The second condition
results from
$\underset{\hspace{-0.3cm}\alpha}{M_{b}^{c}}e_{\beta}^{b}e_{\gamma
c}=0$. And contracting
$\underset{\hspace{-0.3cm}\alpha}{M_{b}^{c}}e_{\beta}^{b}=\lambda
e_{\beta}^{c}$ with $e_{\beta c}$ yields the eigenvalue $\lambda$.
The used simplifications are based on the orthogonality conditions
of the tetrad vectors and on the properties of the rotation
coefficients.

Notice that the two conditions are satisfied if
$n_{\alpha}=n_{\beta}=n_{\gamma}=c$, where $c$ is a constant. The
consequence of this is that $\lambda=0$. In this case,
$k_{ab}=c^{2}\hspace{0.6mm}x_{a}x_{b}+c^{2}\hspace{0.6mm}y_{a}y_{b}+c^{2}\hspace{0.6mm}z_{a}z_{b}$.
\hfill $\square$

The previous theorems show that strong conditions have to be imposed
on $n_{\alpha}$, for $\alpha= 1,2,3$, and the metric in order that
the spatial tetrad vectors are principal directions of
$\underset{\alpha}{M}, \ \ \rm{for} \ \ \alpha= 1,2,3$.

However, the conditions to have $e_{\alpha}$ as eigenvector of
$\underset{\alpha}{M}$ seem less restrictive then the conditions for
$e_{\beta}$, for all values of $ \beta \neq \alpha$, to be
eigenvector of the same tensor $\underset{\alpha}{M}$, since these
involve not only intrinsic derivatives of the scalar fields but also
rotation coefficients of the metric. Furthermore, for $e_{\alpha}$
to be an eigenvector of $\underset{\alpha}{M}$ only conditions on
$n_\alpha$ have to be satisfied, namely that $n_\alpha$ remains
constant along the directions of $e_{\beta}$ for all values of
$\beta \neq \alpha$ (in which case the eigenvalue corresponding to
$e_{\alpha}$ depends only on $n_\alpha$). On the other hand, the
conditions imposed for $e_{\beta}$ for all $ \beta \neq \alpha$ to
be eigenvectors of $\underset{\alpha}{M}$ also involve $n_\beta$ for
all $\beta \neq \alpha$.

Next we will use the previous theorems to establish the conditions
for $e_{\alpha}$, with $\alpha = 1,2,3$ to be an eigenvector of the
three tensors $\underset{1}{M}, \ \underset{2}{M}, \
\underset{3}{M}$ simultaneously, the results being:
\begin{enumerate}
\item[\rm{(i)}] $\Delta_{e_{\beta}}(\ln n_{\alpha})=0 ,$
\item[\rm{(ii)}] $\Delta_{e_{\alpha}}(\ln n_{\beta})=0 ,$
\item[\rm{(iii)}] $\gamma_{\alpha\beta\gamma} \
[n_{\alpha}^{2}-n_{\beta}^{2}]+\gamma_{\alpha\gamma\beta} \
[n_{\gamma}^{2}-n_{\alpha}^{2}]+\gamma_{\beta\gamma\alpha}[n_{\beta}^{2}-n_{\gamma}^{2}]=0,$
\end{enumerate}
for all values of $\beta $ and $\gamma$ such that $\beta \neq \gamma
\neq \alpha$.

These conditions must be satisfied for all values of $\beta \neq
\alpha$. It is not easy to find the general solution to these
equations, however one can say that, in general, the principal
directions of the pulled back material metric $k$ are not, in
general, the principal directions of the three tensors
$\underset{1}{M}$, $\underset{2}{M}$ and $\underset{3}{M}$. It
should be noticed that the (mathematical) solution corresponding to
$n_{1}=n_{2}=n_{3}= \rm{const.}$  is not an interesting result from
the physical point of view.

As a special case, we now consider that all eigenvalues of
$k^{a}_{\hspace{0.15cm}b}$ are equal, i.e. \be
n_{1}=n_{2}=n_{3}=n^{\frac{1}{3}}.\label{lpd} \ee Therefore,
$k_{ab}=n^{\frac{2}{3}}h_{ab}$, so that these tensors are
conformally related. In physical terms, this corresponds to the
unsheared state described in \cite{KS}: the energy per particle,
$\epsilon$, has a minimum under variations of $g^{AB}$ such that $n$
is held fixed. The above theorems in this section simplify
significantly in this case, as can easily be proved using
\eqref{lpd} in those theorems. For completeness, we give the
expressions for the elasticity difference tensor and the tensors
$\underset{\alpha}{M}$ in this special case: \begin{equation*}
S^{a}_{\hspace{0.15cm}bc}=\frac{1}{3}\frac{1}{n}\left(\delta^{a}_{c}D_{b}n+\delta^{a}_{b}D_{c}n-h^{ad}h_{bc}D_{d}n
\right),\end{equation*}
\begin{equation*}
\underset{\hspace{-0.3cm}\alpha}{M_{bc}}=\frac{1}{3}\frac{1}{n}\left(e_{\alpha
c}n_{,b}+e_{\alpha b}n_{,c}+(e_{\alpha c}u_{b}+e_{\alpha
b}u_{c})\Delta_{e_{0}}n-h_{bc}\Delta_{e_{\alpha}}n\right).
\end{equation*}

\section{Examples}

Here, examples concerning the static spherically symmetric case and
an axially symmetric, non-rotating metric are presented, where we
apply the analysis developed in the last section. The main problem
when dealing with examples lies in the difficulties of finding an
orthonormal tetrad for the space-time metric such that the
corresponding spacelike vectors are precisely the principal
directions of the pulled back material metric. However, in the
examples presented, this difficulty was overcome.

\subsection{The static spherically symmetric case}

In this section we analyse the elasticity difference tensor and
corresponding eigendirections for the static spherically symmetric
metric, due to its significance on modelling neutron stars. The
metric regarded here can be thought of
as the interior metric of a non rotating star composed by an elastic material.\\
For a static spherically symmetric spacetime  the line-element can
be written as  \be
ds^{2}=-e^{2\nu(r)}dt^{2}+e^{2\lambda(r)}dr^{2}+r^{2}d\theta^{2}+r^{2}\sin^{2}\theta
d\phi^{2}, \label{dle} \ee where the coordinates
$\omega^{a}=\{t,r,\theta,\phi\}$ are, respectively, the time
coordinate, the radial coordinate, the axial coordinate and the
azimuthal coordinate. Choosing the basis one-forms
$u_{a}=(-e^{\nu(r)},0,0,0)$, $x_{a}=(0,e^{\lambda(r)},0,0)$,
$y_{a}=(0,0,r,0)$ and $z_{a}=(0,0,0,r\sin \theta)$ for the
orthonormal tetrad, the metric is given by $g_{ab}=
-u_au_b+x_ax_b+y_ay_b+z_az_b$ and $h_{ab}=x_ax_b+y_ay_b+z_az_b$
defines the corresponding projection tensor. Using this tetrad, the
pulled-back material metric becomes \be k_{ab}=
n_1^2x_ax_b+n_2^2y_ay_b+n_2^2z_az_b \label{kab}, \ee where we have
chosen $n_3=n_2$ since for this material distribution $k$ has only
two different eigenvalues.

Let $\xi^{A}=\{\tilde{r},\tilde{\theta},\tilde{\phi}\}$ be the
coordinate system in the material space $\mathcal X$. Since the
space-time is static and spherically symmetric, $\tilde{r}$ can only
depend on $r$ and one can take  $\tilde{\theta}=\theta$ and
$\tilde{\phi}=\phi$ so that the configuration of the material is
entirely described by the material radius $\tilde{r}(r)$. Moreover,
the only non-zero components of the deformation gradient are
$\frac{d\xi^{1}}{d\omega^{1}}=\frac{d\tilde{r}}{dr}$,
$\frac{d\xi^{2}}{d\omega^{2}}=1$ and $\frac{d\xi^{3}}{d\omega^{3}}=1$.\\
In $\mathcal X$ the material metric is
$k_{AB}=\tilde{x}_{A}\tilde{x}_{B}+\tilde{y}_{A}\tilde{y}_{B}+\tilde{z}_{A}\tilde{z}_{B}$,
with $\tilde{x}_{A}=e^{\tilde{\lambda}}d\tilde{r}_{A}$,
$\tilde{y}_{A}=\tilde{r}d\tilde{\theta}_{A}$ and
$\tilde{z}_{A}=\tilde{r}sin\tilde{\theta}d\tilde{\phi}_{A}$, and
where ${\tilde{\lambda}} = \lambda(\tilde{r}). $ The pull-back of
the material metric is then  \be
k^{a}_{\hspace{0.15cm}b}=g^{ac}k_{cb}=g^{ac}(\xi^{C}_{c}\xi^{B}_{b}k_{CB})=
\left(\frac{d\tilde{r}}{dr}\right)^{2}e^{2\tilde{\lambda}-2\lambda}\delta^{a}_{1}\delta^{1}_{b}+\frac{\tilde{r}^{2}}{r^{^{2}}}\delta^{a}_{2}\delta^{2}_{b}
+\frac{\tilde{r}^{2}}{r^{2}}\delta^{a}_{3}\delta^{3}_{b}\label{kAB},\ee
Comparing (\ref{kab}) and (\ref{kAB}) it is simple to obtain the
following values for the linear particle densities (all positive),
which are found to depend on $r$ only:
\begin{align}
n_{1}=n_{1}(r)&=\frac{d\tilde{r}}{dr}e^{\tilde{\lambda}-\lambda}
\label{n1}\\
n_{2}=n_{2}(r)&=n_{3}(r)=\frac{\tilde{r}}{r}\label{n2}
\end{align}
The non-zero components of the strain tensor (\ref{2 1}), when
written as functions of the quantities $n_{\alpha}$, are
\begin{align*}
s_{rr}&=\frac{1}{2}e^{2\lambda}(1-n^{-\frac{2}{3}}n_{1}^{2})\\
s_{\theta\theta}&=\frac{1}{2}r^{2}(1-n^{-\frac{2}{3}}n_{2}^{2})\\
s_{\phi\phi}&=\frac{1}{2}r^{2}sin^{2}\theta(1-n^{-\frac{2}{3}}n_{2}^{2})
\end{align*}
Using the expressions obtained for the $n_{\alpha}$ one can find
that the condition for this tensor to vanish identically is that
$\tilde{r}=c e^{\pm \int \frac{1}{r}e^{\lambda -
\tilde{\lambda}}dr}$.

Calculating the quantities given in \eqref{covu} one obtains
\begin{align*}
\Theta&=0\\
\dot{u}_{a}&=\left(0,e^{2\nu}\frac{d\nu}{dr},0,0\right)\\
\sigma_{ab}&:\hspace{0.2cm} \sigma_{12}=\frac{1}{2}e^{4\nu}\frac{d\nu}{dr}=\sigma_{21}\\
\omega_{ab}&:\hspace{0.2cm}
\omega_{12}=e^{2\nu}\frac{d\nu}{dr}+\frac{1}{2}e^{4\nu}\frac{d\nu}{dr}\\
&\hspace{0.5cm}\omega_{21}=-\omega_{12},
\end{align*}
where the remaining components of $\sigma_{ab}$ and $\omega_{ab}$
vanish.

The non-zero components of the elasticity difference tensor
$S^{a}_{bc}$ are:
\begin{align*}
S^{r}_{\hspace{0.15cm}rr}&=\frac{1}{n_{1}}\frac{dn_{1}}{dr}\\
S^{\theta}_{\hspace{0.15cm}\theta r}&=\frac{1}{n_{2}}\frac{dn_{2}}{dr}\\
S^{\phi}_{\hspace{0.15cm}\phi r}&=\frac{1}{n_{2}}\frac{dn_{2}}{dr}\\
S^{r}_{\hspace{0.15cm}\theta\theta}&=r e^{-2\lambda}-r
e^{-2\lambda}\frac{n_{2}^{2}}{n_{1}^{2}}-e^{-2\lambda}r^{2}\frac{n_{2}}{n_{1}^{2}}\frac{dn_{2}}{dr}\\
S^{r}_{\hspace{0.15cm}\phi\phi}&=e^{-2\lambda}r
sin^{2}-e^{-2\lambda}r
sin^{2}\theta\frac{n_{2}^{2}}{n_{1}^{2}}-e^{-2\lambda}r^{2}
sin^{2}\theta\frac{n_{2}}{n_{1}^{2}}\frac{dn_{2}}{dr}.
\end{align*}
Since $S^{a}_{\hspace{0.15cm}bc}=S^{a}_{\hspace{0.15cm}cb}$, there
are only seven non-zero components for this tensor on the coordinate
system chosen above.\\ Again, using (\ref{n1}) and  (\ref{n2})one
obtains that:

(i) the components $S^{\theta}_{\hspace{0.15cm}\theta r}$ and
$S^{\phi}_{\hspace{0.15cm}\phi r}$
are zero whenever the function $\tilde{r}$ is of the form $\tilde{r}=c_{1}r$, where $c_{1}$ is a constant;\\
(ii) $S^{r}_{\hspace{0.15cm}rr}$ is zero whenever
$\tilde{r}=c_{2}+c_{3}\int
e^{\lambda - \tilde{\lambda}}dr;$\\
(iii) the components $S^{r}_{\hspace{0.15cm}\theta\theta}$ and
$S^{r}_{\hspace{0.15cm}\phi\phi}$ are zero whenever
$\tilde{r}=c_{4}e^{\int \frac{e^{-2\tilde{\lambda} +
2\lambda}}{r}dr}$.

The second order symmetric tensors $\underset{\alpha}{M}$, for
$\alpha = 1,2, 3$ have the following non-zero components:
\begin{align*}
\underset{\hspace{-0.3cm}1}{M_{rr}}&=\frac{e^{\lambda}}{n_{1}}\frac{dn_{1}}{dr}\\
\underset{\hspace{-0.3cm}1}{M_{\theta\theta}}&=e^{-\lambda}r-e^{-\lambda}r\frac{n_{2}^{2}}{n_{1}^{2}}-e^{-\lambda}r^{2}\frac{n_{2}}{n_{1}^{2}}\frac{dn_{2}}{dr}\\
\underset{\hspace{-0.3cm}1}{M_{\phi\phi}}&=e^{-\lambda}r
sin^{2}\theta-e^{-\lambda}r
sin^{2}\theta\frac{n_{2}^{2}}{n_{1}^{2}}-e^{-\lambda}r^{2}sin^{2}\theta\frac{n_{2}}{n_{1}^{2}}\frac{dn_{2}}{dr}\\
\underset{\hspace{-0.3cm}2}{M_{r\theta}}&=\underset{\hspace{-0.3cm}2}{M_{\theta
r}}=\frac{r}{n_{2}}\frac{dn_{2}}{dr}\\
\underset{\hspace{-0.3cm}3}{M_{r\phi}}&=\underset{\hspace{-0.3cm}3}{M_{\phi
r}}=\frac{r sin\theta}{n_{2}}\frac{dn_{2}}{dr}
\end{align*}
\noindent The eigenvalues and eigenvectors of these tensors are
presented in tables 1, 2 and 3, being then compared with the
eigendirections of the material metric.

\newpage
\begin{center}
Table 1 - Eigenvectors and eigenvalues for $\underset{1}{M}$
\begin{tabular}{|c|c|}
\hline Eigenvectors & Eigenvalues\\
\hline $x$ & $\mu_{1}=\frac{e^{-\lambda}}{n_{1}}\frac{dn_{1}}{dr}$\\
$y$ &
$\mu_{2}=\frac{e^{-\lambda}}{r}-\frac{e^{-\lambda}}{r}\frac{n_{2}^{2}}{n_{1}^{2}}-e^{-\lambda}\frac{n_{2}}{n_{1}^{2}}\frac{dn_{2}}{dr}$\\
$z$ &
$\mu_{3}=\frac{e^{-\lambda}}{r}-\frac{e^{-\lambda}}{r}\frac{n_{2}^{2}}{n_{1}^{2}}-e^{-\lambda}\frac{n_{2}}{n_{1}^{2}}
\frac{dn_{2}}{dr}$\\
\hline
\end{tabular}
\end{center}
Notice that, in the present example, $\underset{1}{M}$ maintains the
eigenvectors of $k$, namely $x$, $y$ and $z$, the two last ones
being associated with the same eigenvalue. Therefore the canonical
form for $\underset{1}{M}$ is
$\underset{\hspace{-0.3cm}1}{M_{bc}}=\mu_{1}x_{b}x_{c}+\mu_{2}(y_{b}y_{c}+z_{b}z_{c})$,
where $\mu_{1}$ and $\mu_{2}$ are the eigenvalues corresponding to
$x$ and $y$ ($\equiv z$), respectively.
\begin{center}
Table 2 - Eigenvectors and eigenvalues for $\underset{2}{M}$\\
\begin{tabular}{|c|c|}
\hline Eigenvectors & Eigenvalues\\
\hline $x+y$ & $\mu_{4}=\frac{e^{-\lambda}}{n_{2}}\frac{dn_{2}}{dr}$\\
$x-y$ & $\mu_{5}=-\frac{e^{-\lambda}}{n_{2}}\frac{dn_{2}}{dr}$\\
$z$ & $\mu_{6}=0$\\
\hline
\end{tabular}
\end{center}
In this case, only the eigenvector $z$ of $k$ remains as
eigenvector, however the corresponding eigenvalue being zero. The
other two eigenvectors are $x+y$ and $x-y$ so that the canonical
form for $\underset{2}{M}$ can be expressed as
$\underset{\hspace{-0.3cm}2}{M_{bc}}=2\mu_{4}(x_{b}y_{c}+y_{b}x_{c})$,
where
$\mu_{4}=e^{-\lambda}\left(\frac{1}{\tilde{r}}\frac{d\tilde{r}}{dr}-\frac{1}{r}\right)$.
\begin{center}
Table 3 - Eigenvectors and eigenvalues of $\underset{3}{M}$\\
\begin{tabular}{|c|c|}
\hline Eigenvectors & Eigenvalues\\
\hline $x+z$ & $\mu_{7}=\frac{e^{-\lambda}}{n_{2}}\frac{d n_{2}}{dr}$\\
$x-z$ &
$\mu_{8}=-\frac{e^{-\lambda}}{n_{2}}\frac{dn_{2}}{dr}$\\
$y$ &
$\mu_{9}=0$\\
\hline
\end{tabular}
\end{center}
Comparing $\underset{2}{M}$ and $\underset{3}{M}$, it is easy to see
that the role of $z$ and $y$ is interchanged. The eigenvalues of
$\underset{2}{M}$ are equal to the eigenvalues of $\underset{3}{M}$
and the canonical form of this tensor field can be written as
$\underset{\hspace{-0.3cm}3}{M_{bc}}=2\mu_{7}(x_{b}z_{c}+z_{b}x_{c})$,
where
$\mu_{7}=e^{-\lambda}\left(\frac{1}{\tilde{r}}\frac{d\tilde{r}}{dr}-\frac{1}{r}\right)$.

It should be noticed that the case $n_2$ constant is not interesting
to analyze, since this corresponds to the vanishing of the tensors
$\underset{2}{M}$ and $\underset{3}{M}$.

$x$, $y$ and $z$ would only remain eigenvectors for
$\underset{2}{M}$ and $\underset{3}{M}$ if $\tilde{r}$ would be of
the form: $\tilde{r}=c r$, in which case $\underset{2}{M}$ and
$\underset{3}{M}$ were reduced to a zero tensor.

The tetrad components of the elasticity difference tensor can
directly be obtained from \eqref{S2}:
\begin{align*}
S^{1}_{\hspace{0.15cm}11}&=e^{-\lambda}\frac{1}{n_{1}}\frac{dn_{1}}{dr}\\
S^{2}_{\hspace{0.15cm}21}&=e^{-\lambda}\frac{1}{n_{2}}\frac{dn_{2}}{dr}\\
S^{3}_{\hspace{0.15cm}31}&=e^{-\lambda}\frac{1}{n_{2}}\frac{dn_{2}}{dr}\\
S^{1}_{\hspace{0.15cm}22}&= e^{-\lambda}\frac{1}{r}-
e^{-\lambda}\frac{1}{r}\frac{n_{2}^{2}}{n_{1}^{2}}-e^{-\lambda}\frac{n_{2}}{n_{1}^{2}}\frac{dn_{2}}{dr}\\
S^{1}_{\hspace{0.15cm}33}&=e^{-\lambda}\frac{1}{r}
-e^{-\lambda}\frac{1}{r}\frac{n_{2}^{2}}{n_{1}^{2}}-e^{-\lambda}\frac{n_{2}}{n_{1}^{2}}\frac{dn_{2}}{dr}.
\end{align*}

The expressions for the Ricci rotation coefficients are
\begin{align*}
\gamma_{122}&=\frac{e^{-\lambda}}{r}\\
\gamma_{133}&=\frac{e^{-\lambda}}{r}\\
\gamma_{233}&=\frac{\cos\theta}{r \sin\theta}.
\end{align*}

\subsection{The axially symmetric non-rotating case}

First, consider an elastic, axially symmetric, uniformly rotating
body in interaction with its gravitational field. The exterior of
the body may be described by the following metric, \cite{M932}, \be
ds^{2}=-e^{2\nu}dt^{2}+e^{2\mu}dr^{2}+e^{2\mu}dz^{2}+e^{2\psi}(d\phi-\omega
dt)^{2},\label{asr}\ee where $\nu, \psi, \omega, \mu$ are scalar
fields depending on $r$ and $z$.

Assume that the material metric is flat. Introducing in
$\mathcal{X}$ cylindrical coordinates $\xi^{A}=\{R,\zeta,\Phi\}$,
then the material metric takes the form: \be
ds^{2}=dR^{2}+d\zeta^{2}+R^{2}d\Phi^{2}, \ee where the parameters
$R$, $\zeta$ depend on $r$ and $z$, $\Phi$ being
$\Phi(t,r,z,\phi)=\phi-\Omega(r,z)t$.

Now, consider the limiting case of an axially symmetric non-rotating
elastic system for which the space-time metric is given by \be
ds^{2}=-e^{2\nu}dt^{2}+e^{2\mu}dr^{2}+e^{2\mu}dz^{2}+e^{2\psi}d\phi^{2}.\label{asnr}
\ee This metric is obtained from (\ref{asr}), when $\omega=0$ and
the angular velocity $\Omega=0$.

Imposing $R=R(r)$, $\zeta=z$ and $g_{ab}=g_{ab}(r)$, one obtains a
further reduction to cylindrical symmetry. This reduction is
considered in \cite{M932}.

So, the space-time metric we will work with is given by
(\ref{asnr}), where $\nu, \mu, \psi$ depend on $r$ only, and it can
be written as $g_{ab}=-u_{a}u_{b}+x_{a}x_{b}+y_{a}y_{b}+z_{a}z_{b}$,
where $u_{a}=(-e^{\nu(r)},0,0,0)$, $x_{a}=(0,e^{\mu},0,0)$,
$y_{a}=(0,0,e^{\mu(r)},0)$ and $z_{a}=(0,0,0,e^{\psi(r)})$. The
space-time coordinates are $\omega^{a}=\{t,r,z,\phi\}$.

In $\mathcal{X}$ the material metric $k_{AB}$ is given by
$k_{AB}=\tilde{x}_{A}\tilde{x}_{B}+\tilde{y}_{A}\tilde{y}_{B}+\tilde{z}_{A}\tilde{z}_{B}$,
where $\tilde{x}_{A}=dR_{A}$, $\tilde{y}_{A}=dz_{A}$ and
$\tilde{z}_{A}=Rd\phi_{A}$. The relativistic deformation gradient
has the following non-zero components
$\frac{d\xi^{1}}{d\omega^{2}}=\frac{dR}{dr}$,
$\frac{d\xi^{2}}{d\omega^{1}}=1$ and
$\frac{d\xi^{3}}{d\omega^{3}}=1$. Calculating the pull-back of the
material metric one obtains \be
k^{a}_{\hspace{0.15cm}b}=g^{ac}k_{cb}=g^{ac}\left(\xi^{C}_{c}\xi^{B}_{b}k_{CB}\right)=e^{-2\mu}\delta^{a}_{1}\delta^{1}_{b}+\left(\frac{dR}{dr}\right)^{2}e^{-2\mu}\delta^{a}_{2}\delta^{2}_{b}+R^{2}e^{-2\psi}\delta^{a}_{3}\delta^{3}_{b}.
\label{pbmm} \ee The corresponding line-element can be expressed as
\be ds^{2}=dr^{2}+\left(\frac{dR}{dr}\right)dz^{2}+R^{2}d\phi^{2}.
\ee On the other hand, the material metric in the space-time $M$ is
given by \be
k_{ab}=n_{1}^{2}x_{a}x_{b}+n_{2}^{2}y_{a}y_{b}+n_{3}^{2}z_{a}z_{b}.\label{mm}\ee
Comparing (\ref{pbmm}) with (\ref{mm}) one concludes that the linear
particle densities (all positive) are expressed as
\begin{align} n_{1}&=n_{1}(r)=e^{-\mu}\label{n1a}\\
n_{2}&=n_{2}(r)=e^{-\mu}\frac{dR}{dr}\label{n2a}\\
n_{3}&=n_{3}(r)=R e^{-\psi}\label{n3a}. \end{align}

The strain tensor (\ref{2 1}) is composed of the following
components
\begin{align*}
s_{rr}&=\frac{1}{2}e^{2\mu}(1-n^{-\frac{2}{3}}n_{1}^{2})\\
s_{zz}&=\frac{1}{2}e^{2\mu}(1-n^{-\frac{2}{3}}n_{2}^{2})\\
s_{\phi\phi}&=\frac{1}{2}e^{2\psi}(1-n^{-\frac{2}{3}}n_{3}^{2})
\end{align*}
The strain tensor vanishes if the condition $R(r)=r=e^{\psi-\mu}$ is
satisfied.

Calculating the quantities given in \eqref{covu} one obtains
\begin{align*}
\Theta &=0\\
\dot{u}_{a}&=\left(0,e^{2\nu}\frac{d\nu}{dr},0,0\right)\\
\sigma_{ab}&:\hspace{0.2cm} \sigma_{12}=\frac{1}{2}e^{4\nu}\frac{d\nu}{dr}=\sigma_{21}\\
\omega_{ab}&:\hspace{0.2cm}
\omega_{12}=e^{2\nu}\frac{d\nu}{dr}+\frac{1}{2}e^{4\nu}\frac{d\nu}{dr}\\
&\hspace{0.5cm}\omega_{21}=-\omega_{12},
\end{align*}
where the remaining components of $\sigma_{ab}$ and $\omega_{ab}$
vanish.

The non-zero components of the elasticity difference tensor are
\begin{align*} S^{r}_{\hspace{0.15cm}rr}&=\frac{1}{n_{1}}\frac{dn_{1}}{dr}\\
S^{z}_{\hspace{0.15cm}zr}&=\frac{1}{n_{2}}\frac{dn_{2}}{dr}\\
S^{\phi}_{\hspace{0.15cm}\phi
r}&=\frac{1}{n_{3}}\frac{dn_{3}}{dr}\\
S^{r}_{\hspace{0.15cm}zz}&=\frac{d\mu}{dr}-\frac{n_{2}^{2}}{n_{1}^{2}}\frac{d\mu}{dr}-\frac{n_{2}}{n_{1}^{2}}\frac{dn_{2}}{dr}\\
S^{r}_{\hspace{0.15cm}\phi\phi}&=e^{-2\psi-2\mu}\left(\frac{d\psi}{dr}-
\frac{n_{3}^{2}}{n_{1}^{2}}\frac{d\psi}{dr}-\frac{n_{3}}{n_{1}^{2}}\frac{dn_{3}}{dr}\right).
\end{align*} It can be observed that only seven components of the
elasticity difference tensor are non-zero.

Using the expressions (\ref{n1a}), (\ref{n2a}) and (\ref{n3a}) one
can conclude that:

(i) $S^{r}_{\hspace{0.15cm}rr}$ is zero
whenever $\mu(r)=c$, where $c$ is a constant;\\
(ii) $S^{z}_{\hspace{0.15cm}zr}$ is zero whenever
$R(r)=c_{1}+c_{2}\int
e^{\mu(r)}dr$;\\
(iii) $S^{\phi}_{\hspace{0.15cm}\phi r}$ is zero whenever
$R(r)=c_{3}e^{\psi(r)}$;\\
(iv) $S^{r}_{\hspace{0.15cm}zz}$ is zero whenever
$R(r)=\pm\int\sqrt{2\mu(r)+c_{4}}dr+c_{5}$;\\
(v) $S^{r}_{\hspace{0.15cm}\phi\phi}$ is zero whenever
$R(r)=\pm\sqrt{2\int\frac{e^{2\psi}}{e^{2\mu}}\frac{d\psi}{dr}dr+c_{6}}$.

The second-order tensors $\underset{1}{M}$, $\underset{2}{M}$ and
$\underset{3}{M}$ have the following non-zero components:
\begin{align*}
\underset{\hspace{-0.3cm}1}{M_{rr}}&=e^{\mu}\frac{1}{n_{1}}\frac{dn_{1}}{dr}\\
\underset{\hspace{-0.3cm}1}{M_{zz}}&=e^{\mu}\left(\frac{d\mu}{dr}-\frac{n_{2}^{2}}{n_{1}^{2}\frac{d\mu}{dr}}-\frac{n_{2}}{n_{1}^{2}}\frac{dn_{2}}{dr}\right)\\
\underset{\hspace{-0.4cm}1}{M_{\phi\phi}}&=e^{2\psi-\mu}\left(\frac{d\psi}{dr}-\frac{n_{3}^{2}}{n_{1}^{2}}\frac{d\psi}{dr}-\frac{n_{3}}{n_{1}^{2}}\frac{dn_{3}}{dr}\right)\\
\underset{\hspace{-0.3cm}2}{M_{rz}}&=\underset{\hspace{-0.3cm}2}{M_{zr}}=e^{\mu}\frac{1}{n_{2}}\frac{dn_{2}}{dr}\\
\underset{\hspace{-0.3cm}3}{M_{r\phi}}&=\underset{\hspace{-0.3cm}3}{M_{\phi
r}}=e^{\psi}\frac{1}{n_{3}}\frac{dn_{3}}{dr}. \end{align*} The next
three tables contain the eigenvalues and eigenvectors for these
tensors, which are then compared with the eigenvectors of the
pulled-back material metric.
\begin{center}
Table 1 - Eigenvectors and eigenvalues for $\underset{1}{M}$
\begin{tabular}{|c|c|}
\hline Eigenvectors & Eigenvalues\\
\hline $x$ & $\lambda_{1}=e^{-\mu}\frac{1}{n_{1}}\frac{dn_{1}}{dr}$\\
$y$ &
$\lambda_{2}=e^{-\mu}\left(\frac{d\mu}{dr}-\frac{n_{2}^{2}}{n_{1}^{2}}\frac{d\mu}{dr}-\frac{n_{2}}{n_{1}^{2}}\frac{dn_{2}}{dr}\right)$\\
$z$ &
$\lambda_{3}=e^{-\mu}\left(\frac{d\psi}{dr}-\frac{n_{3}^{2}}{n_{1}^{2}}\frac{d\mu}{dr}-\frac{n_{3}}{n_{1}^{2}}\frac{dn_{3}}{dr}\right)$\\
\hline
\end{tabular}
\end{center}
One can observe that the eigendirections $x$, $y$ and $z$ of $k$ are
also eigenvectors for the tensor $\underset{1}{M}$ and the
eigenvectors are associated with different eigenvalues. The
canonical form for $\underset{1}{M}$ can be written as
$\underset{\hspace{-0.3cm}1}{M_{bc}}=\lambda_{1}x_{b}x_c+\lambda_{2}y_by_c+\lambda_{3}z_bz_c$.
\begin{center}
Table 2 - Eigenvectors and eigenvalues for $\underset{2}{M}$
\begin{tabular}{|c|c|}
\hline Eigenvectors & Eigenvalues\\
\hline $x+y$ & $\lambda_{4}=e^{-\mu}\left(\frac{1}{n_{2}}\frac{dn_{2}}{dr}\right)$\\
$x-y$ & $\lambda_{5}=-e^{-\mu}\left(\frac{1}{n_{2}}\frac{dn_{2}}{dr}\right)$\\
$z$ & $\lambda_{6}=0$\\
\hline
\end{tabular}
\end{center}
$\underset{2}{M}$ inherits only the eigenvector $z$ of $k$, which
corresponds to a zero eigenvalue. The other two eigenvectors of
$\underset{2}{M}$ are linear combinations of $x$ and $y$: $x+y$ and
$x-y$, whose corresponding eigenvalues are symmetric in sign. The
canonical form for $\underset{2}{M}$ can be written as
$\underset{\hspace{-0.3cm}2}{M_{bc}}=2\lambda_{4}(x_by_c+y_bx_c)$,
where
$\lambda_{4}=\left(\frac{\frac{d^{2}R}{dr^{2}}}{\frac{dR}{dr}}-\frac{d\mu}{dr}\right)e^{-\mu}$.
\begin{center}
Table 3 - Eigenvectors and eigenvalues for $\underset{3}{M}$
\begin{tabular}{|c|c|}
\hline Eigenvectors & Eigenvalues\\
\hline $x+z$ & $\lambda_{7}=e^{-\mu}\left(\frac{1}{n_{3}}\frac{dn_{3}}{dr}\right)$\\
$x-z$ & $\lambda_{8}=-e^{-\mu}\left(\frac{1}{n_{3}}\frac{dn_{3}}{dr}\right)$\\
$y$ & $\lambda_{9}=0$\\
\hline
\end{tabular}
\end{center}
$\underset{3}{M}$ inherits the eigenvalue $y$ of $k$, which is
associated with the eigenvalue zero. The other two eigenvectors of
$\underset{3}{M}$ are linear combinations of $x$ and $z$: $x+z$ and
$x-z$. These two eigenvectors are associated with sign symmetric
eigenvalues. The canonical form for $\underset{3}{M}$ can be written
as
$\underset{\hspace{-0.3cm}3}{M_{bc}}=2\lambda_{7}(x_{b}z_{c}+z_{b}x_{c})$,
where
$\lambda_{7}=\left(\frac{1}{R}\frac{dR}{dr}-\frac{d\psi}{dr}\right)e^{-\mu}$.

$x$ and $y$ would only be eigenvectors for $\underset{2}{M}$ if
$R(r)$ would be of the form $R(r)=c_{1}+\int e^{\mu}dr c_{2}$, but
in this case $\underset{2}{M}$ would vanish. $x$ and $z$ would only
be eigenvectors for $\underset{3}{M}$ if $R(r)$ would be of the form
$R(r)=c_{3}e^{\psi}$ and this would reduce $\underset{3}{M}$ to a zero tensor.\\
One can observe that the role that $y$ and $n_{2}$ play for the
tensor $\underset{2}{M}$ is the same that $z$ and $n_{3}$ play for
$\underset{3}{M}$. That is, the results for $\underset{2}{M}$ and
$\underset{3}{M}$ are very similar, only $y$ and $n_{2}$ are
substituted by $z$ and $n_{3}$, respectively.

The tetrad components of the elasticity difference tensor obtained
from \eqref{S2} and the expressions for the Ricci rotation
coefficients are listed below:
\begin{align*}
S^{1}_{\hspace{0.15cm}11}&=e^{-\mu}\frac{1}{n_{1}}\frac{dn_{1}}{dr}\\
S^{2}_{\hspace{0.15cm}21}&=e^{-\mu}\frac{1}{n_{2}}\frac{dn_{2}}{dr}\\
S^{3}_{\hspace{0.15cm}31}&=e^{-\mu}\frac{1}{n_{3}}\frac{dn_{3}}{dr}\\
S^{1}_{\hspace{0.15cm}22}&=e^{-\mu}\frac{d\mu}{dr}-e^{-\mu}\frac{n_{2}^{2}}{n_{1}^{2}}\frac{d\mu}{dr}-
e^{-\mu}\frac{n_{2}}{n_{1}^{2}}\frac{dn_{2}}{dr}\\
S^{1}_{\hspace{0.15cm}33}&=e^{-\mu}\frac{d\psi}{dr}-
e^{-\mu}\frac{n_{3}^{2}}{n_{1}^{2}}\frac{d\psi}{dr}-e^{-\mu}\frac{n_{3}}{n_{1}^{2}}\frac{dn_{3}}{dr}.
\end{align*}

The expressions for the Ricci coefficients are
\begin{align*}
\gamma_{122}&=\frac{\frac{d\mu}{dr}}{e^{\mu}}\\
\gamma_{133}&=\frac{\frac{d\psi}{dr}}{e^{\mu}}.
\end{align*}

\section{Acknowledgements}

The authors would like to thank L. Samuelsson for many valuable
discussions on this work.

\end{document}